\documentclass[nofootinbib,prb,twocolumn,showpacs,preprintnumbers,amsmath,amssymb,floatfix,reprint]{revtex4-1}

\usepackage{txfonts}
\usepackage{graphicx}
\usepackage{dcolumn}
\usepackage{bm}
\usepackage{enumerate}

\newcommand{\bfk}{{\bf k}}
\newcommand{\bfr}{{\bf r}}

\newcommand{\req}[1]{\mbox{Eq. ~\!(\ref{#1})}}

\def\G0{G^0}

\def\connect#1{\leavevmode{\setbox1=\hbox{#1}\copy1%
\raise .2\ht1 \vbox{\moveleft \wd1\vbox{\hrule width \wd1 height .5pt depth 0pt}}%
}}

\def\ftn[#1]{\rlap{\footnotemark[#1]}}

\def\H0{H^0}
\def\HMZ{H^0_{\sf M}}

\def\UM{U_{\sf M}}

\def\SHFM{\bar{U}_{\sf M}}

\def\PM{P_{\sf M}}
\def\Pm{P_{\sf m}}
\def\vm{U_{\sf m}}
\def\WM{W_{\sf M}}
\def\Wm{W_{\sf m}}
\def\WD{W_{\sf FP}}
\def\mspace{${\cal M}$}
\def\HM{H_{\sf M}}

\begin{document}

\preprint{APS/123-QED}
\title{Model-mapped RPA for Determining the Effective Coulomb Interaction}

\author{Hirofumi Sakakibara$^{1,2}$}\email{sakakibara.tottori.u@gmail.com}
\author{Seung Woo Jang$^3$}
\author{Hiori Kino$^4$}
\author{Myung Joon Han$^3$} 
\author{Kazuhiko Kuroki$^5$}
\author{Takao Kotani$^1$}

\affiliation{$^1$Department of applied mathematics and physics, Tottori university, Tottori 680-8552, Japan}
\affiliation{$^2$Computational Condensed Matter Physics Laboratory, RIKEN, Wako, Saitama 351-0198, Japan}
\affiliation{$^3$Department of Physics, Korea Advanced Institute of Science and Technology (KAIST), Daejeon 305-701, Korea }     
\affiliation{$^4$Research Organization of Science and Technology, Ritsumeikan university, Kusatsu, Shiga 525-8577, Japan}                                                                                                           
\affiliation{$^4$National Institute for Materials Science (NIMS), Sengen 1-2-1, Tsukuba, Ibaraki 305-0047, Japan}
\affiliation{$^5$Department of Physics, Osaka University, Machikaneyama-Cho, Toyonaka, Osaka 560-0043, Japan}                                                                       
\date{\today}

\begin{abstract}
We present a new method to obtain a model Hamiltonian 
from first-principles calculations. The effective interaction
contained in the model is determined on the basis of random phase approximation (RPA).
In contrast to previous methods such as projected RPA and constrained RPA
(cRPA), the new method named ``model-mapped RPA'' takes into account 
the long-range part of the polarization effect to determine the effective interaction in the model.
After discussing the problems of cRPA,
we present the formulation of the model-mapped RPA, together with
a numerical test for the single-band Hubbard model of HgBa$_2$CuO$_4$.
\end{abstract}


\maketitle

\section{Introduction}
Recently, we often treat low-temperature physical phenomena in correlated
materials by a two-step procedure, that is, deriving a 
model Hamiltonian from a first-principles calculation and solve the
model Hamiltonian
\cite{koretsune-hotta,seo,frg,tohyama}. Thus, the procedure contains two
key points:
\begin{itemize}
 \item[(A)] How to derive a model Hamiltonian.
 \item[(B)] How to solve the model Hamiltonian.
\end{itemize}
For step (B), we can use various many-body calculation techniques to
solve the model Hamiltonian. 
The techniques for step (B) can be dynamical mean-field theory (DMFT)\cite{dmft0},
quantum Monte Carlo (QMC)\cite{QMC} methods, and so on.
The two-step procedure is generally applicable 
to strongly correlated systems such as high-$T_c$
superconductors and metal--insulator transitions.

In this paper, we focus on step (A), that is, 
how to derive a model Hamiltonian from 
first-principles calculations, especially the effective interaction
contained in the Hamiltonian (we neglect phonons here).
If step (A) is well established and combined with a reasonable 
technique in step (B), we can even evaluate the transition
temperature of superconductivity $T_c$
for given crystal structures
without introducing parameters by hand \cite{kent,sakakibara1,sakakibara2,onari,watanabe,sakakibara3,misawa}.
This means we can use the two-step procedure for material informatics
combined with databases of crystal structures. 
In future, we may find new high-$T_c$
superconductors among thousands of possible candidates
\cite{sr2vo4,LaNiO3-LaAlO3} in the two-step procedure.

Let us present an overview of step (A). 
We have various first-principles methods to determine one-body
Hamiltonian $\H0$.
These method are the local density approximation (LDA), the quasiparticle self-consistent GW
method (QSGW)\cite{kotani_quasiparticle_2007,kotani_quasiparticle_2014,di_valentin_spin_2014,klimes_predictive_2014}, 
and so on.
The one-body Hamiltonian $\H0$ describes an independent particle
picture. The static screened Coulomb interaction $W(\bfr,\bfr',\omega=0)$
can be calculated in the random-phase approximation (RPA). 
From $\H0$, we can construct a set of the atomic-like localized orbitals 
$\{\phi_{Ri}(\bfr)\}$ which describe the low-energy bands.
The orbitals can be constructed, for example, by the method of
the maximally localized Wannier functions 
\cite{marzari_maximally_1997}. 
The orbitals span a model Hilbert space \mspace. 
The choice of \mspace\ is not unique and has ambiguity.
If necessary, we should use a larger \mspace\ to reduce the ambiguity.
However, step (B) requires a sufficiently small \mspace\
to ensure tractability within current computational resources.
The one-body part of the model Hamiltonian in \mspace
can be determined by the projection of $\H0$ into \mspace. 
As for the effective interaction, 
we cannot simply project $W(\bfr,\bfr',\omega=0)$ into \mspace. 
In advance, we have to remove the screening effect
expected within the model. This is necessary to avoid double counting of the screening effect.
This idea was first introduced in the projected RPA (pRPA) method by
Kotani \cite{kotani_ab_2000}. This was followed by the constrained RPA (cRPA) by 
Aryasetiawan {\it et al.} \cite{aryasetiawan_frequency-dependent_2004}.
cRPA has been applied to several compounds to analyze  
strongly correlated systems
\cite{nakamura-arita,nakamura-nohara,nakamura-yoshimoto,nomura2,nomura1,tsuchiizu}. 
Miyake implemented a Wannier-based modified cRPA in the {\tt ecalj} package
\cite{miyake_ab_2009}. 
\c{S}a\c{s}{\i}o\v{g}lu, Friedrich, and Bl\"{u}gel 
also proposed a modified cRPA applicable to cases with entangled bands
 \cite{sasioglu_effective_2011}.
However, cRPA contain theoretical problems as we discuss in Sec. \ref{revcrpa}.

The reliability of the model Hamiltonian obtained in step (A) is 
determined by the reliability of the first-principles calculation.
Most popular calculations
are in LDA. 
However, LDA often gives an unreliable independent particle picture,
especially for transition metal oxides and $f$-electron materials. 
A well-known problem is the underestimation of the band gaps.
Furthermore, there are problems with the bandwidth, the positions of the 
3$d$ bands and 4$f$ bands relative to the oxygen bands, and so on. 
In such cases, we need to use advanced methods 
such as hybrid functional methods \cite{kim_towards_2010} 
or QSGW
\cite{kotani_quasiparticle_2007,kotani_quasiparticle_2014,di_valentin_spin_2014,klimes_predictive_2014}.
One of the advantages of QSGW is that $\H0$ and $W(\bfr,\bfr',\omega=0)$
are determined simultaneously in a self-consistent manner.
QSGW has even been applied to metallic ground states.
For example, Han {\it et al.} recently applied QSGW to LaNiO$_3$/LaAlO$_3$\cite{han_quasiparticle_2014}, 
Jang {\it et al.} applied QSGW to high-$T_c$ materials 
\cite{jang_quasiparticle_2015}, and Ryee {\it et al.} applied QSGW to SrRuO$_3$ 
and Sr$_2$RuO$_4$ \cite{ryee2016a}. To handle such metallic
systems, QSGW is more reliable than the hybrid functional methods \cite{heyd_hybrid_2003}.
Furthermore, Deguchi, Sato, Kino, and Kotani have recently shown that 
a QSGW-based hybrid method can systematically give a good description 
for a wide range of materials \cite{deguchi2016}.
These calculations were performed by the first-principles calculation package {\tt ecalj} \cite{ecalj},
which is based on a mixed-basis full-potential method, 
the linearized augmented plane wave, and muffin-tin orbital
method (the PMT method) \cite{kotani_formulation_2015,kotani_linearized_2013,kotani_fusion_2010}.
It is freely available from github \cite{ecalj}.

In this paper, we propose a new method named model-mapped RPA (mRPA).
This is based on an assumption of the 
existence of a model Hamiltonian that explains the low-energy physical properties of materials.
This assumption is standard in the field of model calculations.
For example, we may assume that low-energy 
physical properties can be quantitatively understood
by a Hubbard model. Then, the role of mRPA is to
determine the interaction parameters in the model.

After we review cRPA and point out its problems in Sec. \ref{revcrpa},
we give a formulation of mRPA in Sec. \ref{fmrpa}.
Then we show how it works for a test case of single-band Hubbard model
for the high-$T_c$ superconductor HgBa$_2$CuO$_4$
in Sec. \ref{numtest}, followed by a summary.

\section{cRPA and its problems}  
\label{revcrpa}
In the first-principles calculations, 
the screened Coulomb interaction 
$W(\bfr,\bfr',\omega)$ in RPA is given by
\begin{eqnarray}
W=\frac{1}{1-vP}v,
\label{eq:w}
\end{eqnarray}
where $v(\bfr,\bfr')$ and $P(\bfr,\bfr',\omega)$ are the Coulomb
interaction and the non-interacting proper polarization, respectively.
$\frac{1}{X}$ (written as $X^{-1}$ below) denotes the inverse of matrix $X$.
$P$ consists of a product of two Green functions $G_0=1/(\omega-\H0)$.
We can represent the quantities $W,v,$ and $P$ 
expanded in an improved version of the mixed product basis (MPB).
The MPB was originally introduced by Kotani in 
Ref. \onlinecite{kotani02}. Then, the MPB was 
improved by Friedrich, Bl\"{u}gel, and Schindlmayr \cite{friedrich_efficient_2010}. 
We usually use the improved MPB.

Let us recall the idea of the so-called cRPA.
We first choose a model space \mspace\ spanned by a basis set of 
atomic-like localized orbitals,
$\{\phi_{Ri}(\bfr)\}$, where $R$ is the index of the primitive cell and $i$
is the index used to specify an orbital in the cell.
In the following, we use the notations, $1\equiv n_1 \equiv R_1 i_1$ and
\begin{eqnarray}
 &&(1,2|W|2',1') = (\phi_{1},\phi_{2}|W|\phi_{2'},\phi_{1'}) \nonumber \\
 &\equiv& \int d^3r d^3r' 
 \phi^*_1(\bfr) \phi^*_2(\bfr') W(\bfr,\bfr',\omega) \phi_{2'}(\bfr') \phi_{1'}(\bfr),
\end{eqnarray}
as in the manner of Ref. \onlinecite{lichtenstein_ab_1998}.
The eigenfunctions in \mspace\ 
are calculated from the Hamiltonian 
$\HMZ \equiv
\langle \phi_{Ri}|\H0|\phi_{R'i'}\rangle$, that is,
$\HMZ$ is the same as $\H0$ but restricted within the space \mspace.
Then, we have the non-interacting proper polarization function
$\Pm(\bfr,\bfr',\omega)$ in \mspace.
Let us consider the RPA-screened Coulomb
interaction $\Wm(\bfr,\bfr',\omega)$  in \mspace.
It can be written as
\begin{eqnarray}
\Wm=\frac{1}{1-\vm\Pm}\vm,
\label{eq:Wm}
\end{eqnarray}
where $\vm(\bfr,\bfr',\omega)$ is the (not yet determined) effective interaction 
between quasi-particles in \mspace.
cRPA determines $\vm(\bfr,\bfr',\omega)$ by assuming
\begin{eqnarray}
\Wm=W,
\label{eq:wwm}
\end{eqnarray}
with \req{eq:Wm}.
From Eqs. (\ref{eq:w}), (\ref{eq:Wm}), and (\ref{eq:wwm}), we have
\begin{eqnarray}
\vm=\frac{1}{1-v(P-\Pm)}v,
\label{eq:vm}
\end{eqnarray}
that is, we can calculate $\vm$ from $v,P,$ and $\Pm$.
Then, we calculate the on-site interaction $U$ for the model as
\begin{eqnarray}
U\equiv U_{n,n,n,n}=(\phi_{n},\phi_{n}|\vm|\phi_{n},\phi_{n}).
\label{eq:URn}
\end{eqnarray}
If necessary, we can calculate any elements of the interaction given as
$U_{1,2,2',1'}=(\phi_1,\phi_2|\vm|\phi_{2'},\phi_{1'}$).
However, in the usual cRPA, we only calculate 
the set of parameters $\{U\}$ used by the model.
In summary, for the choice of a localized basis set $\{\phi_{Ri}\}$,
we determine a set of interactions $\{U\}$ of the model in cRPA,
where $U$ are $\omega$-dependent.

The cRPA, which appears to be reasonable, however,
contains the following three problems.
\begin{itemize}
\item[(i)]{\bf Range truncation problem}

$W=\Wm$ is satisfied only when we take all possible elements of $U_{1,2,2',1'}$ in cRPA.
However, practically adopted models consider a limited number of $U$.
Note that $\vm$ given in \req{eq:vm} is inevitably long-range.
This is because the strong screening effects such as metallic screening
contained in $\Pm$ are removed from the total polarization $P$.
This problem is well illustrated when only the on-site $U$ is used.
In this case, because we use only the on-site part of $\vm$ 
evaluated from the right-hand side of \req{eq:vm}, $\Wm$ given in \req{eq:Wm}
cannot satisfy the condition $\Wm=W$.

Sch\"uler {\it et al.} \cite{schuler_optimal_2013} proposed a method 
to solve an extended Hubbard model with non local interaction. 
However, the method is not applicable to the long-range
interaction $\propto 1/|\bfr-\bfr'|$ without modification. 
Hansmann {\it et al.} \cite{hansmann_long-range_2013} calculated
the long-range behavior of the effective interaction $\propto
1/|\bfr-\bfr'|$ by cRPA, 
and presented a method to solve a model Hamiltonian taking into account
the long-range interaction. In contrast, we can
handle the same problem using a model Hamiltonian with the short-range
interaction given by mRPA. This is
because mRPA downfolds the long-range interaction 
into the short-range interaction as described in Sec. \ref{fmrpa}.

\item[(ii)] {\bf Positive definiteness and causality problem}

$-(P-\Pm)$ in the denominator of \req{eq:vm}
should be positive definite at $\omega=0$. 
If this is not satisfied, we obtain unphysical results for 
$\vm$ having eigenvalues larger than the bare interactions $v$.
Furthermore, the imaginary (anti-hermitian) part of
$-(P-\Pm$) at any $\omega$ should be positive definite so as to 
satisfy causality. In the original idea of cRPA, $P-\Pm$ does not 
necessarily satisfy this condition for the case of entangled bands.
For this reason, Kotani avoided the idea of cRPA 
and proposed pRPA, which satisfies the above conditions \cite{kotani_ab_2000}.
Recently, two other procedures satisfying the
conditions have been proposed 
with a modification of the definition of $P-\Pm$ in the cRPA.
One given by Miyake, Aryasetiawan, and Imada,
neglects the off-diagonal elements between \mspace\ and residual space in
the one-body Hamiltonian [see Fig. 1 and Eq. (8) in Ref. \onlinecite{miyake_ab_2009}].
Thus, the condition is clearly satisfied.
The other is given by \c{S}a\c{s}{\i}o\v{g}lu, Friedrich, and Bl\"ugel,
where a projection procedure of eigenfunctions to \mspace\ is used to satisfy the condition \cite{sasioglu_effective_2011}.
Strictly speaking, neither procedure should be identified as cRPA,
since the key advantage of cRPA, \req{eq:wwm}, is no longer satisfied.

Note the generality of the causality problem. For example, in the GW+DMFT 
formulation \cite{biermann_first-principles_2003,sun} as 
an extension of the LDA+DMFT \cite{dmft1,dmft2,dmft3}, 
the on-site part of the GW self-energy is simply substituted with 
the DMFT self-energy. Then, we may have a causality problem
if we have a GW self-energy whose imaginary part is
larger than that of the DMFT self-energy.

\item[(iii)] {\bf energy window problem}\\
In Table \ref{table:nipara}, we have calculated the static (at $\omega=0$) part of $W$ and $U$ for the paramagnetic
Ni, where we use the cRPA method given by \c{S}a\c{s}{\i}o\v{g}lu, Friedrich, and Bl\"ugel
\cite{sasioglu_dftu_2014}. 
We considered two cases for the energy window; the narrower one 
is $-8\sim 1$ eV and the wider is $-10\sim 10$ eV.
In contrast to the small difference in $W$ for the different energy
window, we see a large difference in $U$. 
The value of 3.56 eV is in good agreement with that in Ref. \onlinecite{sasioglu_effective_2011}. 
As shown in Fig. 5 in Ref. \onlinecite{sasioglu_dftu_2014}, they used such a wide energy window.
The value of 2.25 eV for the narrower window is significantly different from 3.56 eV.
This difference is because of the difference in $\Pm$, which describes
the polarization of the 3$d$ electrons.
In the case of wider windows, we remove more polarization, 
resulting in larger values of $U$. This results in an inevitable ambiguity in the cRPA 
because we have almost the same energy bands 
(the same eigenvalue dispersions in the Brillouin zone) for both windows.
In addition, we have no definite criteria for choosing a certain energy window.\\
We expect that a similar ambiguity also exists in other versions of cRPA.
Miyake, Aryasetiawan, and Imada,
successfully obtained flat low-energy behaviors, as shown in Fig. 3 
of Ref. \onlinecite{miyake_ab_2009}, similar to that in Fig. 1 of Ref. \onlinecite{kotani_ab_2000}.
However, the procedure of neglecting the off-diagonal elements
(equivalent to how to choose the \mspace) is ambiguous.
From Fig. 3 in Ref. \onlinecite{miyake_ab_2009}, 
we guess that the ambiguity of $U(\omega=0)$ in their method 
can be $\simeq$ 1 eV [from the degree of freedom in the choice of the energy window and 
\mspace, we may have various possible extrapolations of $U(\omega)$ to $\omega=0$].

\end{itemize}

\begin{table}[t]
\caption{Static screened Coulomb interactions $W$ and $U$ 
in cRPA method \cite{sasioglu_effective_2011}
for two different outer energy windows, 
-8$\sim $1 eV, and -10$\sim$ 10 eV for paramagnetic Ni. 
For both energy windows, we have almost the same 3$d$ bands
for the model space \mspace.
We use 12 $\times$ 12 $\times$ 12 $\bfk$ points in the Brillouin zone in the tetrahedron
method \cite{kotani_quasiparticle_2007,kotani_quasiparticle_2014}.
There is a large difference between the two $U$ values. See text.}
\label{table:nipara}
\begin{tabular}{l| c c c c}
\hline
       & -8$\sim$ 1 eV  & -10$\sim$ 10 eV \\
\hline
$W$[eV]  &  1.19  & 1.40  \\
$U$[eV]  &  2.25  & 3.56  \\
\hline
\end{tabular}
\end{table}

Although the new method, the mRPA, formulated in Sec. \ref{fmrpa}
can remedy these problems, we need to pay attention to the inevitable limitations 
of model Hamiltonians including no long-range interactions.
Recall that plasmons (charge fluctuations) do not satisfy the Goldstone's
theorem because of the $1/r$ behavior of the Coulomb interaction. 
Such model Hamiltonians cannot describe this correctly.
The long-range limit of longitudinal spin fluctuations, as well.
Model Hamiltonians can only be justified when these problems are irrelevant.

\section{Formulation of the mRPA}
\label{fmrpa}
Let us assume that a model Hamiltonian $\HM$ in the model space \mspace\ can describe 
low-energy excitations very well. Here we formulate mRPA, which determines
the parameters included in $\HM$. $\HM$ is given as
\begin{eqnarray}
\HM= \HMZ + \UM - \SHFM, 
\label{eq:hm}
\end{eqnarray}
where $\HMZ$ is the one-body Hamiltonian, obtained from a
first-principles method such as QSGW.
$\UM$ is the spin-independent effective interaction specified by
a set of parameters $\{U\}$ 
(here we do not consider the $\omega$-dependence of these parameters).
The terms $\HMZ + \UM$ are those in Eq. (1) in Ref. \onlinecite{lichtenstein_ab_1998}. 
$\SHFM$ is the one-body
counter term so that the effect of $\UM-\SHFM$ is
canceled out when we apply the first-principles method 
to the model described by $\HM$ \cite{ikeda-arita}.
In the $\UM$, used elements $\UM(1,2,2',1')$ 
are given by a set of a finite number of parameters $\{U\}$.
We usually allow only the short-range terms; for example, 
we only allow the on-site terms in the case of the Hubbard model.

Let us explain how to determine $\{U\}$ (or $\UM$ equivalently) in 
mRPA. If we apply RPA to the model Hamiltonian $\HM$,
we have the screened Coulomb interaction of the model $\WM(1,2,2',1')$ as
\begin{eqnarray}
\WM = \frac{1}{1-\UM \PM} \UM,
\label{eq:WM}
\end{eqnarray}
where we use the proper non-interacting polarization $\PM$ calculated from $\HMZ$.
In mRPA, we only consider the case at $\omega=0$ in Eqs. (\ref{eq:WM})--(\ref{eq:UM2}). 

Note the difference between \req{eq:Wm}(cRPA) and \req{eq:WM}(mRPA).
In \req{eq:Wm}, $\vm$ inevitably become long-range as $\propto 1/r$,
while $\UM$ is short-range such as on-site only in Hubbard model.
That is, $\PM,\UM$ and $\WM$ are non-zero just on the limited number 
of discrete index set of \mspace\ in \req{eq:WM}.

For the theoretical correspondence, we require $\WM$ to satisfy 
\begin{eqnarray}
\WM(1,2,2',1') = \WD(1,2,2',1'),
\label{eq:WMW}
\end{eqnarray}
in mRPA in order to determine $\UM$. 
Here, $\WD(1,2,2',1')\equiv(1,2|W|2',1')$ is the quantity calculated from $W$ 
 in the first-principles method using \req{eq:w}.
It is not possible to satisfy \req{eq:WMW} for all the matrix
elements of $\WM(1,2,2',1')$; we satisfy a subset of \req{eq:WMW}
corresponding to the degree of freedom of $U(1,2,2',1')$ used in $\UM$
of \req{eq:hm}.
Thus, the subset of \req{eq:WMW} can determine $\{U\}$ uniquely.
Then, we can determine $\UM$ from \req{eq:WM} so as to satisfy
\begin{eqnarray}
\frac{1}{1-\UM \PM} \UM = \WD,
\label{eq:UM}
\end{eqnarray}
or 
\begin{eqnarray}
\UM = \frac{1}{1+ \WD \PM}{\WD}, 
\label{eq:UM2}
\end{eqnarray}
equivalently. 

By definition, mRPA satisfies \req{eq:WMW} exactly,
where $\WM(1,2,2',1')$ is expressed in terms of $\UM$ and $\PM$ in \req{eq:WM}.
This is in contrast to the case of cRPA, which can not usually satisfy
\req{eq:wwm} since cRPA usually discards the off-site part of $\vm$.
Thus, we are free from the problem (i) in Sec. \ref{revcrpa} in the case
of mRPA.

We may have cases that  
$\UM$ satisfying \req{eq:WMW} cannot be found.
This is because $\WM$ has the upper limit $-1/\PM$ 
($\approx$ the bandwidth of $\HMZ$);
for $\UM\to \infty$, we have $-1/\PM$ can be seen from \req{eq:WM}.
Thus, we cannot determine $\UM$ for very large $\WD$. This can be
clearly seen in Fig. \ref{fig:WM} as explained later.
In such cases, we need to use a larger \mspace.
This is not an intrinsic problem of mRPA but a problem
associated with choosing of too small \mspace. 

The causality problem (ii) in Sec. \ref{revcrpa} does not arise,
since $\omega$-dependence of $\UM$ is meaningless in mRPA;
we use the condition given by \req{eq:WMW} only at $\omega=0$. 
In our opinion, we rather have to use a larger \mspace\ for better results,
instead of taking the $\omega$-dependence into account 
in theoretical treatments. 
If we take the $\omega$-dependence
of the effective interaction correctly, 
we inevitably have to treat a quantum Langevin equation with electron thermal bath.
Such a treatment is far beyond our current numerical techniques because 
it requires an enormous computational effort. 

Let us consider problem (iii) in Sec. \ref{revcrpa} in the case of mRPA.
In mRPA, $\PM$ is only determined from the energy bands of $\HMZ$.
The choice of the Wannier functions (the choice of \mspace)
can slightly change $\WD$. This yields the slight ambiguity of $\UM$ via
\req{eq:UM}.
This is inevitable as long as we derive a model from first-principles
calculations. In contrast, cRPA has further ambiguity 
in the polarization of $\Pm$ in \req{eq:vm} owing to 
the ambiguity of the choice of Wannier functions as we have shown in
Table \ref{table:nipara}.

\section{Numerical Test for a Single Band Hubbard Model of HgBa$_2$CuO$_4$}
\label{numtest}
Here we present a test calculation to see how mRPA works in
comparison with cRPA.
We take a single-band Hubbard model for stoichiometric HgBa$_2$CuO$_4$.
We treat two cases where $\HMZ$ is determined by LDA or by QSGW.
The space \mspace\ is chosen by a procedure based on maximally localized
Wannier functions \cite{souza_maximally_2001}.
The term $\SHFM$ in \req{eq:hm} is irrelevant in 
the single-band case since it gives a constant potential shift.
As we use the single-band Hubbard model, 
we obtain $\UM$ and $\WM$ as scalars.

\begin{figure}[t]
\includegraphics[width=9cm]{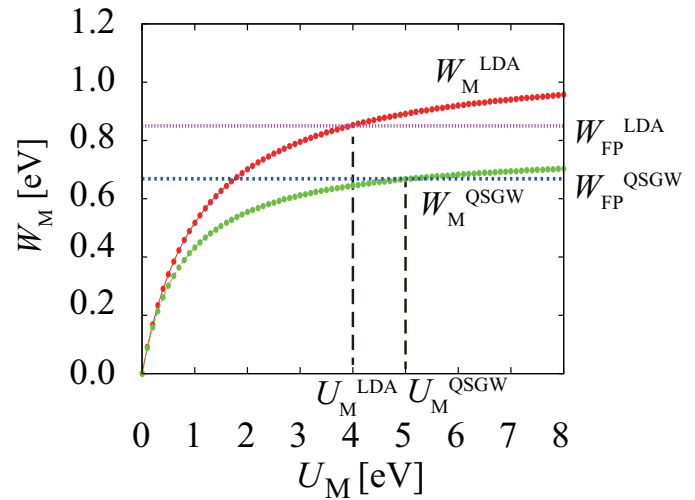}
\caption{(Color online) Calculated $W_{\sf M}$ as a function of $\UM$ in \req{eq:WM} 
 for a single-band model of HgBa$_2$CuO$_4$.
The Green line represents QSGW, the read line represents LDA.
The values of $\UM$ in mRPA are obtained at the intersections 
between the curves of $\WM$ and the horizontal lines of $\WD$.
\label{fig:WM}
} 
\end{figure}

The two curves in Fig. \ref{fig:WM} show $\WM$ as functions of $\UM$
given by \req{eq:WM}, where we use $\PM$ calculated from $\HMZ$ by QSGW
or by LDA. As a function of $\UM$, these curves are initially linear near $\UM=0$ and
saturate toward $\frac{-1}{\PM}$. The difference between the two
curves is due to the size of $-1/\PM$ corresponding to the size of the bandwidth \cite{jang_quasiparticle_2015}.
The two horizontal lines show the values of $\WD$ (0.67 and
0.85 eV) calculated by the first-principles RPA method as shown in Table \ref{table:UW}.
Using the condition \req{eq:WMW}, we can determine $\UM$ for QSGW and for LDA
as illustrated in Fig. \ref{fig:WM}. 

The obtained values of $\UM$ are shown in Table
\ref{table:UW}, together with the cRPA values $\vm$ 
obtained by the method in Ref. \onlinecite{sasioglu_effective_2011}.
The values of $\UM$ are $1.5\sim2.0$ eV larger than those of $\vm$.
This is because  we use the on-site interaction only in the present model.
If we take into account off-site interactions, $\UM$ will be
reduced.
In other words, mRPA downfolds the off-site interactions into the on-site interaction.

\begin{table}[t]
\label{table1}
\caption{
Calculated values of $\UM$ (mRPA) and $\vm$ (cRPA) for a single-band model of HgBa$_2$CuO$_4$ 
, together with $\WD$.
Note that we only consider the values at $\omega=0$.
We use the tetrahedron method in Ref. \onlinecite{kotani_quasiparticle_2007}
for the evaluation of $\UM$ and $\WD$, where
we use 8 $\times$ 8 $\times$ 4 $\bfk$ points in the Brillouin zone, as was used
in Ref. \onlinecite{werner_dynamical_2015}.
The values of $\vm$ in cRPA 
\cite{sasioglu_effective_2011}
are the same as those presented in our previous paper \cite{swj-crpa}.
$\UM$ is determined by mRPA as illustrated in Fig. \ref{fig:WM}. }
\begin{tabular}{l| c c c c} 
\hline
      & $\WD$ [eV]    & $\UM$ [eV] & $\vm$ [eV]  \\  \hline 
QSGW  &  0.67      &    5.2    & 3.7                   \\
LDA   &	 0.85    &       3.9      & 2.0            \\
\hline
\end{tabular}
\label{table:UW}
\end{table}

In Fig. \ref{fig:WM}, we see that the determined values of $\UM$ are sensitive 
to the values of $W_{\rm FP}^{\rm LDA}$ and $W_{\rm FP}^{\rm QSGW}$.
This is because the calculated values of $\UM$ are close to the 
upper limit of RPA, $-1/\PM$ at $\UM\to \infty$.
The derivatives $dW/dU$ at $\UM$ are rather small, 
0.060 for LDA and 0.019 for QSGW.
This sensitivity may indicate the lack of suitability (or limitation) of the single-band Hubbard
models for HgBa$_2$CuO$_4$. If we use a larger \mspace, we will be able to
avoid such cases of $\UM \sim -1/\PM$.

\section{Summary}
We have presented mRPA to determine model Hamiltonians
based on first-principles calculations.
mRPA is formulated starting from the assumption of the existence of a model Hamiltonian
that explains the low-energy physical properties of materials. Then we 
determine the effective interactions contained in the Hamiltonian by matching
the first-principles RPA calculations and the RPA calculations using
the model Hamiltonian.
mRPA is free from the theoretical problems in cRPA, which are
discussed in Sec. \ref{revcrpa}. Thus, mRPA is less ambiguous and
logically clearer than cRPA.
Through the model Hamiltonian obtained by mRPA, 
we will be able to predict the critical temperatures of superconductors
based on first-principles calculations.

We appreciate discussions with Drs. Friedlich, \c{S}a\c{s}{\i}o\v{g}lu, Miyake, and Arita. 
H.S. appreciates fruitful discussions with Yunoki, Shirakawa, Seki, and Shinaoka.
This work was supported by JSPS KAKENHI (Grant-in-Aid for Young Scientists B, Grant No. 16J21175),
and was partly supported by the Advanced Low Carbon Technology Research and Development Program (ALCA) of Japan Science and Technology Agency (JST).
S.W.J. and M.J.H were supported by the Basic Science Research Program
through NRF (2014R1A1A2057202).  The computing resource were
supported by KISTI (KSC-2015-C3-042) , the supercomputing system Great-Wave (HOKUSAI) of RIKEN, 
the supercomputing system of the ISSP, and 
the Computing System for Research in Kyushu University. 

\bibliographystyle{apsrev4-1}

\bibliography{condmat-mrpa.bib}

\end{document}